\documentclass[prb,twocolumn,showpacs,preprintnumbers,amsmath,amssymb]{revtex4}
\usepackage{amsfonts}
\usepackage{graphicx}
\usepackage{dcolumn}
\usepackage{float}
\usepackage{bm}
\usepackage{color}

\renewcommand{\O}{ {\cal{O}} }

\newcommand{\com}[1]{}

\begin{document}

\title{Bloch-Redfield theory of high-temperature magnetic fluctuations in interacting spin systems}

\author{Andrew Sykes}

\affiliation{Theoretical Division and Center for Nonlinear Studies, Los Alamos National Laboratory, Los Alamos, NM 87545, USA}

\author{Dmitry Solenov}

\affiliation{Naval Research Laboratory, Washington, D.C. 20735, USA}

\author{Dmitry Mozyrsky}\email{mozyrsky@lanl.gov}

\affiliation{Theoretical Division (T-4), Los Alamos National Laboratory, Los
Alamos, NM 87545, USA}

\date{\today}

\begin{abstract}

We study magnetic fluctuations in a system of interacting spins on a lattice at high temperatures and in the
presence of {a} spatially varying magnetic field. Starting from a microscopic Hamiltonian we derive effective
equations of motion for the spins and solve these equations self-consistently. 
We find that the spin fluctuations can be described by an effective diffusion
equation with {a} diffusion coefficient {which strongly depends} on the ratio of the magnetic field gradient to the strength
of spin-spin interactions. We also extend our studies to account for external noise and 
find that the relaxation
times {and} the diffusion coefficient are mutually dependent.

\end{abstract}

\maketitle
\section{Introduction}

Recent advances in magnetic imaging techniques{,} as 
well as {the} development of novel types of electronic devices 
that utilize electronic spin (rather than charge) as an 
information carrier{,} have renewed interest in understanding 
mechanisms of spin noise and spin relaxation. While conventional 
experimental methods, such as nuclear or electron spin 
resonance and related techniques \cite{amb, Schlihter}, 
probe {the} temporal evolution of spin correlations, they typically 
do not provide much information on spatial correlations between 
neighboring spins. On the contrary, the new approaches to spin 
resonance, such as magnetic resonance force microscopy (MRFM), combine 
capabilities of the usual magnetic resonance techniques with {the} sensitivity 
of atomic force microscopy. That is, one can now observe not 
only the time (frequency) dependence of spin correlations, 
but also their spatial dispersion with an atomic-scale resolution.  
Hence{,} there is a clear need to develop theoretical tools for the 
description of such correlations in systems of interest, that is, 
in systems of {\it interacting} spins.

The spatial correlations in interacting spin systems 
are believed to be controlled by the so-called {\it flip-flop} 
processes. That is, two neighboring interacting spins can exchange 
magnetization, i.e., the values of their spin components can change by 
$\pm 1/2$, so that the total spin of the pair is conserved. Such 
exchange gives rise to the diffusion of spin magnetization, 
provided the dynamics of the flip-flops is Poissonian \cite{anderson}. 
Typical calculations of the effective diffusion constant utilize 
the method of moments, where the line-width is approximated by a 
gaussian or lorentzian shape\cite{Schlihter}{. Such} 
approximations are not very well controlled. More recently several 
types of cluster/cummulant expansions have been proposed in connection 
with the problem of decoherence of localized electronic spins caused 
by the fluctuations of nuclear spins \cite{souza, vitzel, saikin}. In 
that problem though, the decoherence of electronic spins occurs on a timescale 
small compared to the typical nuclear timescale, which justifies 
the use of cluster expansions in the description of fluctuations in the nuclear subsystem.

\begin{figure}
\includegraphics[width=8cm]{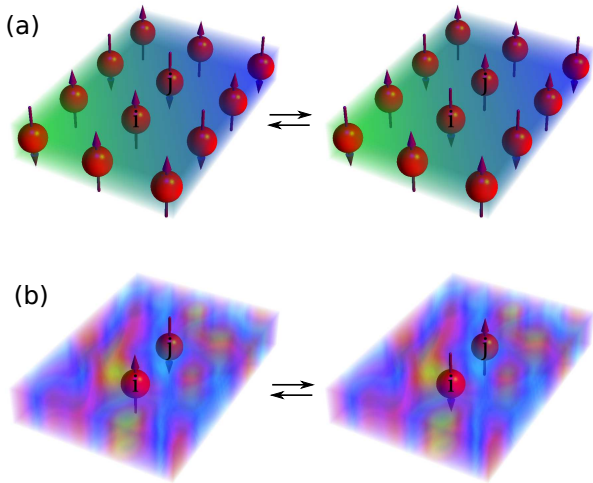}
\caption{(a) shows a collection of spin-half particles on a rigid lattice in a nonuniform external magnetic field. The quantity of
interest in this work is the rate at which spin flip-flops occur. Our model is displayed pictorially in (b), where the
neighbouring sites of $i$ and $j$ are replaced by a fluctuating bath. The flip-flop rate is then calculated such that
it is consistent with the fluctuations of the bath.}\label{fig3}
\end{figure}

In this paper we study correlations between spatially separated spins in 
the opposite, long time regime. Such {a} regime is specifically relevant 
to the MRFM technique, which utilizes (micro)mechanical cantilevers 
with ferromagnetic tips to probe magnetic fluctuations in the underlying 
samples. We propose an approach based on {the} Markov approximation, similar to 
the frequently used Bloch-Redfield approximation\cite{Schlihter, blum} in 
the theory of open quantum systems. That is, we consider all possible pairs 
$(i,j)$ of interacting spins, while other spins $\neq(i,j)$ are treated as 
{an} environment, providing finite line-width for the flip-flop transitions through 
fluctuating magnetic fields {(see Fig.~\ref{fig3} for a cartoon visualisation of 
these approximations)}. A self-consistency is then established between 
the flip-flop rates and the line-width so that our approach can be viewed 
as a sort of dynamical mean field approximation. We argue that our method 
is well justified, in particular, in the presence of an external strongly 
non-uniform magnetic field, which introduces separation between the 
timescales of the flip-flop rates and the correlation time for the fluctuations 
of the effective magnetic fields. Note that such non-uniform magnetic fields 
are intrinsic to the MRFM setups, where field gradients are used to address 
specific spins located within the so-called resonance layer.

Our paper is organized as follows. In {Section~\ref{sec:model}} we describe a 
general formalism that can be utilized to study spin-spin correlations 
for a broad class of spin Hamiltonians, e.g. Eq. (\ref{ham}). We derive 
effective equations of motion for the magnetization, e.g. Eq. (\ref{sigz3}), 
which has the form of a stochastic master equation. In doing so we use 
methods developed in connection with studies of diffusion in classical 
lattice gas models \cite{sasha, kogan} as well as in the theory of open 
quantum systems \cite{blum}. The equation of motion is supplemented by 
a self-consistency equation, Eq. (\ref{Gamma}), which relates the 
rates in the master equation to the correlation function evaluated 
from the master equation in terms of the rates. In Section {\ref{heisenberg} 
we look specifically at }
the Heisenberg model on a cubic lattice in the presence of a spatially 
non-uniform external magnetic field. We find that the {flip-flop} rates are 
strongly suppressed by the field gradient in the limit when the field 
gradient significantly exceeds the spin-spin interaction constant. 
In Section \ref{relaxation} we study the influence of spin-relaxation 
processes on spin flip-flops and derive the effective master equation 
for the magnetization in the presence of external noise sources acting on 
the spins. Our main result of that section is {that,} while the field gradient 
suppresses the flip-flops, the noise may actually enhance these rates; 
see Eq. (\ref{lorentzian}) and corresponding discussion. Finally, in 
{Section \ref{sec:discussion}} we discuss the validity of our approximations 
and summarize the results.

\section{Model and general solution}\label{sec:model}

We consider a system of spin-half particles on a lattice, interacting with 
each other according to the following Hamiltonian
\begin{equation}\label{ham}
H\!=\!\sum_i B_i \sigma^z_i+\sum_{\langle i,j\rangle}\left[J_{ij}^\parallel\sigma_i^z\sigma_j^z+
2J_{ij}^\perp\left(\sigma_i^+\sigma_j^-+\sigma_i^-\sigma_j^+\right)\right]
\end{equation}
where $\sigma^{\pm}_k=(\sigma^x_k\pm i\sigma^y_k)/2$, $k=(k_x,k_y,k_z)$, 
and $\sigma^\alpha_k$ are Pauli matrices, $\alpha=x, y, z$.
The index $i$ in the first sum runs over all lattice sites, while the 
notation $\langle i,j\rangle$ in the second sum indicates the
summation over all pairs of lattice sites.
The external magnetic field $B_i$ is assumed to be non-uniform in space. 
The spin-spin interaction is isotropic
when $J^\parallel_{ik}=J^\perp_{ik}$. The equation of motion for $\sigma^z_i$ is
\begin{equation}\label{eq:sigmaz}
i\partial_t\sigma^z_i= [\sigma^z_i, H]=4\sum_{{k\neq i}} J^\perp_{ik}
(\sigma^+_i\sigma^-_k - \sigma^-_i\sigma^+_k).
\end{equation}
In Eq. (\ref{eq:sigmaz}) and in the following we set $\hbar = 1$. Next we consider {the} equation of
motion for $\sigma^+_i\sigma^-_k$. After a straightforward calculation we obtain
\begin{align}\label{eq:sigmapm}
i\partial_t(\sigma^+_i\sigma^-_k)=& 2\,\delta\!B_{ki}^{\rm eff}\sigma^+_i\sigma^-_k + J^\perp_{ik}(\sigma^z_i-\sigma^z_k)
\\\nonumber
&+2\!\!\sum_{ n\neq \{i,k\}}\left[ J^\perp_{ni}\sigma^z_i \sigma^+_n\sigma^-_k - J^\perp_{nk}\sigma^z_k \sigma^+_i\sigma^-_n\right],
\end{align}
where $\delta\!B_{ki}^{\rm eff}$ is the difference between effective magnetic fields at sites $k$ and $i$,
\begin{equation}\label{Beff}
\delta\!B_{ki}^{\rm eff}=B_k-B_i+\sum_{n\neq \{k,i\}}\left[J^\parallel_{nk}\sigma_n^z-J^\parallel_{ni}\sigma_n^z\right].
\end{equation}
This difference consists of a constant part;
\begin{equation}\label{Bconst}
 \delta\!B_{ki}=B_k-B_i,
\end{equation}
and a part which fluctuates (due to spin flips at nearby lattice sites);
\begin{equation}\label{Bfluct}
\delta\!B_{ki}^{\rm fluct}(t)=\sum_{n\neq \{k,i\}}\left[J^\parallel_{nk}\sigma_n^z-J^\parallel_{ni}\sigma_n^z\right].
\end{equation}
The larger the number of individual spins contributing to $\delta\!B_{ki}^{\rm fluct}$, the more rapidly fluctuating this
quantity becomes. Hence, for systems with sufficiently long-range interactions or high dimensionality
$\delta\!B_{ki}^{\rm fluct}$ fluctuates very rapidly.

From Eq. (\ref{eq:sigmapm}) we see that the expectation value of $\sigma^+_i\sigma^-_k$ contains a prefactor
\begin{equation}\label{Delta}
\Delta_{ik}(t,s)=e^{2i\int_s^t dt^\prime \,\delta\!B_{ik}^{\rm eff}(t^\prime)},
\end{equation}
related to {the} Larmor precession of spins around the effective magnetic field at sites $i$ and $k$. The fluctuating
component of the effective-magnetic-field [see Eq.~\eqref{Bfluct}] causes the precession frequencies at each site to vary.
Moreover, if the effective magnetic fields at sites $i$ and $k$ are large, and the number of spins contributing to the
fluctuating component of the field [see Eq. (\ref{Bfluct})] is much greater than one, then from Eq.~\eqref{eq:sigmapm}
[or more specifically, {the prefactor shown in} Eq.~\eqref{Delta}], we would expect
the Larmor precession frequency of $\sigma^+_i\sigma^-_k$ to be very fast (compared to the dynamics of the individual
$\sigma^z_n$ operators) and fluctuate rapidly. Following this logic, we see that the summation of terms
$\sigma_n^+\sigma_k^-$ and $\sigma_i^+\sigma_n^-$ in Eq. (\ref{eq:sigmapm}) is essentially a summation over a rapidly
fluctuating object, and will statistically self-average to zero (provided a sufficiently large number of spins contribute to
$\delta\!B_{ki}^{\rm fluct}$).

A similar approximation is very common in the theory of
open quantum systems, where it is known as the secular or Bloch-Redfield approximation \cite{blum}. As in the case of
open quantum systems it relies on the assumption that the off-diagonal elements of a system's density matrix $\rho$ are
small either due to large splittings between the adjacent energy levels or due to rapid fluctuations from the heat bath.
In the present case the fields $\delta\!B_{ki}^{\rm fluct}(t)$, play the role of the heat bath operators and must
treated self-consistently, to which we now focus our attention.

By integrating Eq. (\ref{eq:sigmapm}) (with the summation on the {right-hand-side} neglected) 
we obtain
\begin{eqnarray}\label{eq:sigmasol}
\sigma^+_i\sigma^-_k(t)\simeq -i J^\perp_{ik}\int_0^t ds \Delta_{ik}(t,s)[\sigma^z_i(s)-\sigma^z_k(s)]
\\\nonumber
+\Delta_{ik}(t,0){c_{ik}},
\end{eqnarray}
where the last term is due to the initial condition of the operator {$c_{ik}=\sigma^+_i\sigma^-_k(t=0)$}. In the high temperature limit
the system is disordered and therefore it is natural to assume that the expectation value of $\sigma^+_i\sigma^-_k$ is
random, with $\langle\langle \sigma^+_i\sigma^-_k \rangle\rangle=0$ and
$\langle\langle \sigma^+_i\sigma^-_k \sigma^+_{i^\prime}\sigma^-_{k^\prime}\rangle\rangle=(1/4)\delta_{ik^\prime}\delta_{ki^\prime}$,
provided $i\neq k$ and $i'\neq k'$.
Here the double bracket stands
for averaging over the ensemble of density matrices of the system 
as well as over a particular realization of the
density matrix {(set by a particular choice of the initial condition)}, i.e., $\langle \sigma^+_i\sigma^-_k \rangle= {\rm Tr}(\sigma^+_i\sigma^-_k\rho)$ and
$\langle\langle \sigma^+_i\sigma^-_k \rangle\rangle = \langle{\rm Tr}(\sigma^+_i\sigma^-_k\rho)\rangle_\rho$, etc.

We wish to substitute Eq. (\ref{eq:sigmasol}) into  Eq. (\ref{eq:sigmaz}) to obtain a closed form equation for $\sigma^z_i(t)$.
This can be significantly simplified if we replace the rapidly fluctuating quantity,
$\Delta_{ik}(t,s)$, in the integrand in Eq. (\ref{eq:sigmasol}) by
its average value. This approximation is in a perfect agreement with our assumption regarding the separation between time scales for
the dynamics of the local fluctuating magnetic field at site $i$, and components of the individual
spin at site $i$. We make the assumption that, by virtue of the central limit theorem, the random variable
$\delta\!B_{ik}^{\rm eff}$ is Gaussian;
\begin{align}
 \langle\Delta_{ik}(t,s)\rangle&=e^{2i(B_i-B_k)(t-s)}e^{-2\int_s^t\int_s^tK_{ik}(\tau_1-\tau_2)d\tau_1d\tau_2}\nonumber\\
&=e^{2i(B_i-B_k)(t-s)}e^{-4\int_0^{|t-s|}K_{ik}(\mu)\left(|t-s|-\mu\right)d\mu}\label{AvDelta}
\end{align}
where
\begin{equation}\label{K}
 K_{ik}(\tau_1-\tau_2)=\langle\delta\!B_{ik}^{\rm fluct}(\tau_1)\delta\!B_{ik}^{\rm fluct}(\tau_2)\rangle
\end{equation}
is the autocorrelation function of the fluctuating component of the magnetic field gradient between sites $i$ and $k$.
Moreover, since Eq.~\eqref{AvDelta} (as a function of $|t-s|$) decays much faster than the evolution of
$[\sigma^z_i(s)-\sigma^z_k(s)]$, we can employ the Markov approximation, and set $s=t$ which removes
the latter term from the integral in Eq.~\eqref{eq:sigmasol} to give
\begin{eqnarray}\label{eq:sigsol2}
\sigma^+_i\sigma^-_k(t)\simeq -i J^\perp_{ik}\int_0^t ds \langle\Delta_{ik}(t,s)
\rangle[\sigma^z_i(t)-\sigma^z_k(t)]
\\\nonumber
+\Delta_{ik}(t,0){c_{ik}}.
\end{eqnarray}
Now that we have a formal solution for $\sigma^+_i\sigma^-_k(t)$ it is prudent to substitute the
expression back into the sum in Eq.~\eqref{eq:sigmapm} which was originally ignored in deriving
Eq.~\eqref{eq:sigsol2}. In doing so we wish to find an inequality which quantitatively ensures
the summation term is small compared to all other terms in Eq.~\eqref{eq:sigmapm}. The details of
this calculation are straightforward (see {Section}~\ref{sec:discussion} for further discussion)
and one finds
 $J_{ik}\ll\Gamma_{ik}$ {(where $\Gamma_{ik}$ is the rate at which flip-flops occur
and is calculated below)}
is a sufficient condition to ensure the summation in Eq.~\eqref{eq:sigmapm} remains small.

We now substitute Eq.~\eqref{eq:sigsol2} into Eq.~\eqref{eq:sigmaz}, to give
\begin{align}\label{sigz3}
 \partial_t\langle\sigma_k^z(t)\rangle=&\sum_{j\neq k}\Gamma_{jk}\left[\langle\sigma_j^z(t)\rangle-\langle\sigma_k^z(t)\rangle\right]+\xi_k(t).
\end{align}
The averages in Eq. (\ref{sigz3}) are taken with respect to a particular
realization of {the} systems density matrix, but not over the ensemble of the density matrices.
The coefficient, $\Gamma_{jk}$, represents a rate at which spin flip{-flops} occur between sites $j$ and $k$
({these} can only occur when sites $j$ and $k$ have opposite spin). The expression for this
rate is given by
\begin{align}
 \Gamma_{jk}=&4(J_{jk}^\perp)^2\int_{-\infty}^{\infty}\exp\!\!\Bigg[2i\delta\! B_{jk}s-\nonumber\\
&\qquad
\left.4\int_0^{|s|}K_{kj}(\mu)\left(|s|-\mu\right)d\mu\right]ds,\label{Gamma}
\end{align}
where we have used the quickly-decaying property of $\langle\Delta_{ik}(t,s)\rangle$ to extend the
upper and lower limits of the integral to $\pm\infty$. The final term in Eq.~\eqref{sigz3}
represents the uncertainty with respect to the choice of the initial conditions of the system, and is given by
\begin{equation}\label{xi}
 {\xi_k(t)=4i\sum_{j\neq k}J_{jk}^\perp\left[c_{jk}\Delta_{kj}(0,t)-c_{kj}\Delta_{jk}(0,t)\right].}
\end{equation}
Averaging over $\xi_i(t)$ corresponds to averaging over {an ensemble of different density matrices 
(each density matrix being distinguished by a unique initial condition)}.
Noting that $\langle\Delta_{ik}(0,t)\Delta_{ki}(0,t')\rangle=\langle\Delta_{ki}(t',t)\rangle$, and
since $\Delta_{ik}(t',t)$ is a rapidly fluctuating function of $t-t'$, we can make the approximation;
\begin{equation}\label{fluct}
\langle\xi_i(t)\xi_{j}(t^\prime)\rangle = 2\delta(t-t')\left(-\Gamma_{jk}+\delta_{jk}\sum_{m\neq k}\Gamma_{mk}\right).
\end{equation}

Together, {Eqs.} (\ref{sigz3}) and (\ref{fluct}) obviously describe Poissonian dynamics of {a} coupled
two-state system. Indeed, we could have obtained the same result if we had postulated that the
dynamics of a given spin (say, at site $i$) is controlled by its flipping rates
$-\sum_k \tilde{\Gamma}_{ik} \sigma_i(1-\sigma_k)$ and $\sum_k \tilde\Gamma_{ki} \sigma_k(1-\sigma_i)$,
where $\tilde{\Gamma}_{ik}=\Gamma_{ik}+\eta_{ik}$, with $\Gamma_{ik}$ and $\eta_{ik}$ being the constant
and fluctuating parts of the rate respectively. In this case $\xi_i=\sum_k (\eta_{ik}-\eta_{ki})$, c.f. Eq.~(\ref{xi}).
Note that one can derive Eq.~(\ref{Gamma}) for the rates $\Gamma_{ik}$ within a 
straightforward perturbative
calculation{, as shown in Appendix~\ref{appA}. There, we calculate the probability 
of a flip-flop for a pair of spins in
the presence of an external fluctuating field (along {the $z$}-direction). 
In the current section, we have simply assumed that this fluctuating external field 
has been created by the neighbouring spins coupled to 
this pair (see Appendix~\ref{appA} for details).}

Equations (\ref{sigz3}) and (\ref{fluct}) 
constitute a closed system of equations, which allows one
to evaluate the correlation functions $\langle\sigma^z_i(t)\sigma^z_k(t^\prime)\rangle$. For an arbitrary
choice of spin-spin interaction constants $J^\parallel_{ik}$ and $J^\perp_{ik}$ and external fields $B_i$,
the rates $\Gamma_{ik}$ in Eqs. (\ref{sigz3}) and (\ref{fluct}), though formally unkown, are expressed in terms of
these correlation
functions $\langle\sigma^z_i(t)\sigma^z_k(t^\prime)\rangle$ [see Eqs. \eqref{Gamma}, \eqref{K}, 
and \eqref{Bfluct}].
{By} evaluating these correlation functions in terms of $\Gamma_{ik}$, one obtains {a closed
set of} equations which one must solve self consistently for $\Gamma_{ik}$.
This provides a way of solving for both the rates, $\Gamma_{ik}$ and the correlation functions,
$\langle\sigma^z_i(t)\sigma^z_k(t^\prime)\rangle$ for an arbitrary choice of
interaction constants; $J^\parallel_{ik}$, $J^\perp_{ik}$ and external fields; $B_i$.

In the next section we will evaluate the
$\Gamma_{ik}$ and $\langle\sigma^z_i(t)\sigma^z_k(t^\prime)\rangle$ for 
a simple choice of coupling constants
given by the three dimensional, cubic, Heisenberg model with nearest-neighbor interactions.

Before proceeding to this task we note that in the limit of large field gradient
$|B_i-B_k|\gg J^\parallel_{ik}$, the integrand {of} Eq. (\ref{Gamma}) rapidly
oscillates and therefore the value of the integral decreases with the 
growth of $|B_i-B_k|$. In the limit
of vanishing rate $\langle\sigma^z_i(t)\sigma^z_k(t^\prime)\rangle\simeq \delta_{ik}${, 
we find}
\begin{equation}\nonumber
K_{ik}(t-t^\prime)\simeq \kappa_{ik}= \sum_{m\neq\{i,k\}}(J_{mk}^\parallel-J_{mi}^\parallel)^2.
\end{equation}
{Evaluating then, }the Gaussian integral in Eq. (\ref{Gamma}) we obtain
\begin{equation}\label{Gamma1}
\Gamma_{ik}\simeq \frac{4\pi^{1/2}( J^\perp_{ik})^2}{\sqrt{2 \kappa_{ik}}}\exp{\!\left[-\frac{\delta\! B_{ik}^2}{2 \kappa_{ik}}\right]}.
\end{equation}
Thus we predict the rate at which flip-flops occur, {and therefore the rate at which 
spin diffusion occurs,} is very small for $|B_i-B_k|\gg J^\parallel_{ik}$.

\section{Example: Heisenberg model}\label{heisenberg}

We now consider a particular example; {the} Heisenberg model on a cubic lattice with an external spatially varying magnetic field.
The Hamiltonian of the system can be cast in the form
\begin{equation}\label{ham1}
H=\sum_i B({\bf r}_i)\sigma^z_{{\bf r}_i}+ J \sum_{i, \nu, \alpha}\sigma^\alpha_{{\bf r}_i}\,\sigma^\alpha_{{\bf r}_i+{\bf e}_\nu}.
\end{equation}
where $i=(i_x, i_y, i_z)$, ${\bf r}_i=i_x a {\hat {\bf x}}+i_y a {\hat {\bf y}}+i_z a {\hat {\bf z}}$ ($a$ being the lattice spacing),
$\nu = 1, ..., 6$ enumerates the unit vectors which point to the nearest neighbors:
${\bf e}_{1(2)}=\pm\hat {{\bf x}}$, ${\bf e}_{3(4)}=\pm\hat {{\bf y}}$ and ${\bf e}_{5(6)}=\pm\hat {{\bf z}}$,
and finally $\alpha=x,y,z$.
We also assume that the external field varies linearly in space, $B({\bf r})= b_0 {\bf r}\cdot {\bf g}$ where ${\bf g}$ is a
unit vector which points in the direction of variation. The Hamiltonian (\ref{ham1}) obviously belongs
to the class of Hamiltonians defined in Eq. (\ref{ham}).

The equation of motion for $\sigma^z_{{\bf r}_i}$ is given by Eq. (\ref{sigz3}), which, for the Hamiltonian in Eq. (\ref{ham1}) reads
\begin{equation}\label{ham2}
\partial_t\langle\sigma^z_{{\bf r}_i}\rangle =\sum_\nu\Gamma_{\bf{e}_\nu}\left[\langle\sigma^z_{{\bf r}_i+{\bf e}_\nu}\rangle-\langle\sigma^z_{{\bf r}_i}\rangle\right] +\xi_{{\bf r}_i}(t)
\end{equation}
and the noise $\xi_{{\bf r}_i}(t)$ is correlated according to Eq. (\ref{fluct}), which becomes
\begin{equation}\label{noise2}
\langle\xi_{{\bf r}_i}(t)\xi_{{\bf r}_j}(t^\prime)\rangle = 2\delta(t-t^\prime)\sum_\nu
\Gamma_{\bf{e}_\nu} (\delta_{{\bf r}_i\,{\bf r}_j}-\delta_{{\bf r}_i+{\bf e}_\nu \,{\bf r}_j}).
\end{equation}

Eqs. (\ref{ham2}) and (\ref{noise2}) can be readily diagonalized by a Fourier 
transform method. Writing
\begin{equation}\label{fourier}
\sigma^z_{{\bf r}_i}(t) = \int_{-\infty}^{\infty}\frac{dt}{2\pi} \int_{-\pi/a}^{\pi/a}\frac{d^3{\bf k}}{(2\pi)^3}
{\tilde\sigma}^z({\bf k}, \omega) e^{i{\bf r}_i {\bf k}+i\omega t},
\end{equation}
where the ${\bf k}$-integral is taken over the first Brillouin zone, 
(a cube with an edge $2\pi/a$), we obtain from Eq. (\ref{ham2}) that 
\begin{equation}\label{fourier2}
\langle |{\tilde\sigma}^z({\bf k},\omega)|^2\rangle =
\frac{\langle |{\tilde\xi}({\bf k},\omega)|^2\rangle}{ \omega^2 + 
\{\sum_\nu \Gamma_{\bf{e}_\nu}\left[1-\cos{(a\,{\bf e}_\nu {\bf k})}\right]\}^2},
\end{equation}
with $i=x, y, z$ and ${\tilde\xi}({\bf k}, \omega)$ {being the Fourier transform of 
$\xi_{{\bf r}_i}(t)$, defined 
similarly} to Eq. (\ref{fourier}). From Eq. (\ref{noise2})
\begin{equation}\label{noise2f}
\langle |{\tilde\xi}({\bf k},\omega)|^2\rangle = 2\sum_\nu \Gamma_{\bf{e}_\nu}\left[1-\cos{(a\,{\bf e}_\nu {\bf k})}\right],
\end{equation}
and taking the inverse Fourier transform of Eq. (\ref{noise2f}), we obtain
\begin{align}
\langle \sigma^z_{{\bf r}_i}(t) \sigma^z_{{\bf r}_{i^\prime}}(0) \rangle = \,&e^{-t\sum_\nu \Gamma_{\bf{e}_\nu}}\times\nonumber
\\
&\qquad I_{n_x}(2\Gamma_{{\bf e}_1} t)I_{n_y}(2\Gamma_{{\bf e}_3} t)I_{n_z}(2\Gamma_{{\bf e}_5} t),\label{corr}
\end{align}
where $I_n(z)$ is {the} modified Bessel function of complex argument \cite{grad} and
$n_x = |i_x-i_x^\prime|$, etc. At sufficiently large distances (and times) Eq. 
(\ref{corr}) describes (anisotropic) diffusion with diffusion constants $D_{\nu\nu}\sim \Gamma_{\bf{e}_\nu} a^2$.

\begin{figure}
\includegraphics[width=8cm]{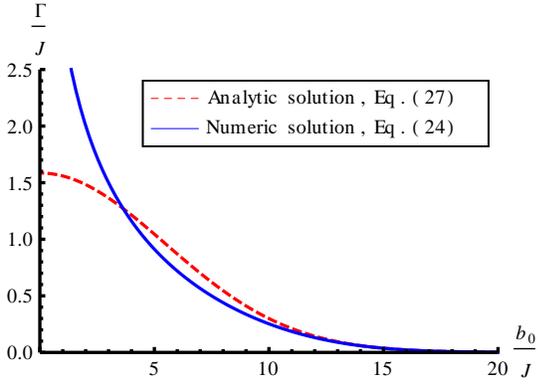}
\caption{Numerical solution of the integral Eq.~\eqref{Gamma2}, showing the rate $\Gamma$ as a function
of magnetic field gradient $b_0$.}\label{fig1}
\end{figure}

The rates $\Gamma_{\bf{e}_\nu}$ are yet to be determined. They can be found from Eq. (\ref{Gamma}).
Note that while for arbitrary direction of the field gradient ${\bf g}$ the rates $\Gamma_{\bf{e}_\nu}$, 
$\Gamma_{\bf{e}_{\nu'}}$ 
differ from each other, they are equal ($\Gamma_{\bf{e}_\nu}\equiv\Gamma$) for
${\bf g}={\bf g}_0=(1/\sqrt{3})({\hat {\bf x}}+{\hat {\bf y}}+{\hat {\bf z}})$, i.e., when the field
gradient points along the main diagonal of the {cube} formed by the unit vectors $\hat {{\bf x}}$, $\hat {{\bf y}}$
and $\hat {{\bf z}}$. In this case Eq. (\ref{fluct}) reduces to
\begin{equation}\label{Gamma2}
\Gamma = 4J^2\int_{-\infty}^\infty ds \,e^{2ib_0 s/\sqrt{3}}e^{-4\int_0^{|s|}d\mu K(\mu)\left(|s|-\mu\right)},
\end{equation}
where $K(\mu)$ is the correlation function of Eq.~\eqref{K}, which is now independent of the indices $i$ and $k$, due
to our convenient choice of magnetic field gradient direction ${\bf g}$, which makes the diffusion process isotropic.
$K(\mu)$ can be easily expressed in terms of $\langle \sigma^z_{{\bf r}_i}({\mu}) \sigma^z_{{\bf r}_j}(0)\rangle$:
\begin{align}\nonumber
K({\mu}) = \,&\frac{1}{2}J^2 \left(\sum_{\nu\neq1}\sum_{\nu^\prime\neq1}\langle\sigma^z_{{\bf e}_\nu}({\mu})
\sigma^z_{{\bf e}_{\nu^\prime}}(0)\rangle\right.
\\&\left.\label{corr2}
-\sum_{\nu\neq1}\sum_{\nu^\prime\neq2}\langle\sigma^z_{{\bf e}_\nu}({\mu}) \sigma^z_{{\bf e}_1+{\bf e}_{\nu^\prime}}(0)\rangle\right),
\end{align}
where we have chosen to calculate $K$ between sites ${\bf r}_i=(0,0,0)$ and ${\bf r}_j={\bf e}_1$ (and then relied on the
isoptropy of all directions in the lattice).
Using Eq. (\ref{corr}) we obtain $K(\mu) = \frac{1}{2}J^2 e^{-6\Gamma \mu}f(2\Gamma \mu)$, where
\begin{align}
f(x) = &5I_0^3(x)+16 I_0(x) I_1^2(x)+ 4 I_0^2(x) I_2(x) \nonumber
\\\nonumber
&- 4I_0^2(x) I_1(x)-8I_1^3(x)- 12 I_0(x)I_1(x)I_2(x) \nonumber
\\
&- I_0^2(x)I_3(x).\label{f}
\end{align}
Substituting this new found expression for $K(\mu)$ into Eq. (\ref{Gamma2}) we obtain an integral equation for $\Gamma$.
One can solve this integral equation numerically to find
$\Gamma/J$ as a function of $b_0/J$ (see Appendix~\ref{appB}), the results are shown in Fig.~\ref{fig1}.
For $b_0/J\gg 1$ the value of $\Gamma$ is
consistent with Eq. (\ref{Gamma1}), which for the present case reduces to
\begin{equation}\label{Gamma3}
{\Gamma \simeq 4J\sqrt{\frac{\pi}{20}}e^{-b_0^2/(60J^2)}.}
\end{equation}
{The analytic solution is also shown in Fig~\ref{fig1} for comparison. We find 
the analytic and numerical solutions are equal beyond $b_0\gtrsim 10J$.}

\section{Influence of relaxation processes}\label{relaxation}
In this section we consider the influence of external noise on {the} spin-spin correlation function. 
We consider a model described by {the} Hamiltonian
\begin{equation}\label{hamnoise}
{\tilde H} = H + \sum_{i,\alpha} \eta^\alpha_i(t)\,\sigma^\alpha_i,
\end{equation}
where $H$ is given by Eq. (\ref{ham}) and $\eta^\alpha_i(t)$ is a 
fluctuating magnetic field. The index $i$ runs over lattice sites, and
$\alpha=x,y,z$. In {reality} such a field may 
arise due to phonons ({for instance in semiconductors}) or conduction electrons ({for instance in metals}). We will assume that
$\langle\eta^\alpha_i(t) \eta^{\beta}_{j}(t^\prime)\rangle = \delta_{\alpha\beta}\,\delta_{ij}\,\Lambda(t-t^\prime)$, where $\Lambda(t)$ is some
even function which decays to zero over some time scale.

We follow a similar {procedure} as in Section \ref{sec:model}. By calculating commutation relations{,} we find;
\begin{equation}
 i\partial_t\sigma_k^z=4\sum_{j\neq k}J_{jk}^\perp\left(\sigma_k^+\sigma_j^--\sigma_j^+\sigma_k^-\right)+
4\left(\eta_k^-\sigma_k^+-\eta_k^+\sigma_k^-\right)\label{sigz}
\end{equation}
where $\eta_k^\pm=\frac{1}{2}\left(\eta_k^x\pm i\eta_k^y\right)$.
\begin{equation}
 i\partial_t\sigma_k^+=-2B_k^{\rm eff}\sigma_k^++2\eta_k^+\sigma_k^z+2\sum_{j\neq k}J_{jk}^\perp\sigma_k^z\sigma_j^+
\label{sig+}
\end{equation}
where $B_k^{\rm eff}=B_k+\sum_{j\neq k}J^\parallel_{jk}\sigma_j^z+\eta_k^z$ is the effective magnetic field at site $k$. Also
\begin{equation}\label{sig-}
 i\partial_t\sigma_k^-=2B_k^{\rm eff}\sigma_k^--2\eta_k^-\sigma_k^z-2\sum_{j\neq k}J_{jk}^\perp\sigma_k^z\sigma_j^-.
\end{equation}
Finally,
\begin{align}\nonumber
 i\partial_t\left(\sigma_j^+\sigma_k^-\right)=\,&2\Delta\!B_{kj}^{\rm eff}\sigma_j^+\sigma_k^-+J_{jk}^\perp\left(\sigma_j^z-\sigma_k^z\right)
+\\&\nonumber
2\left[\eta_j^+\sigma_j^z\sigma_k^--\eta_k^-\sigma_j^+\sigma_k^z\right]
+\\&\label{sig+sig-}
2\sum_{i\neq\{j,k\}}\left[J_{ij}^\perp\sigma_j^z\sigma_i^+\sigma_k^--J_{ik}^\perp\sigma_k^z\sigma_j^+\sigma_i^-\right]
\end{align}
where $\Delta\!B_{kj}^{\rm eff}=\delta\!B_{kj}^{\rm eff}+\eta_k^z-\eta_j^z$,
and $j\neq k$. Analagous to Eqs.~\eqref{Bconst} and~\eqref{Bfluct} of Section~\ref{sec:model}, $\Delta\!B_{kj}^{\rm eff}$ consists
of a constant part, given by ${\delta\!B_{kj}}$ [see Eq.~\eqref{Bconst}], and a fluctuating part, which is now given by
\begin{equation}
 \Delta\!B_{kj}^{\rm fluct}(t)=\delta\!B_{kj}^{\rm fluct}(t)+\eta_k^z(t)-\eta_j^z(t),
\end{equation}
{compared with Eq.~\eqref{Bfluct}.} 
We wish to integrate Eqs. \eqref{sig+}, \eqref{sig-}, and \eqref{sig+sig-}, and thereby find a closed form for the time evolution
of $\sigma_k^z$ from Eq. \eqref{sigz}.

We start with Eqs. \eqref{sig+}, and \eqref{sig-} and apply the same logic as in Section \ref{sec:model} regarding the self-averaging
nature of the summations (due to a fluctuating Larmor precession frequency). What is left can easily be integrated to give
\begin{align}
 \sigma_k^\pm(t)=\,&\mp2i\int_0^t \left[e^{\pm2i\int_s^t B_k^{\rm eff}(\tau)d\tau}\eta_k^\pm(s)\sigma_k^z(s)\right]ds+\nonumber\\
&\quad c_k^\pm e^{\pm2i\int_0^t B_k^{\rm eff}(\tau)d\tau},
\end{align}
where $c_k^\pm=\sigma_k^\pm({t=}0)$ gives the contribution from the initial conditions.
Looking now at Eq. \eqref{sig+sig-}{,} and ignoring the summation term{,} we find
\begin{align}
 \sigma_j^+\sigma_k^-(t)\!=\!-i\int_0^t\Delta_{jk}'(s,t)
\Bigg\{\left[J_{jk}^\perp+2\eta_j^+(s)\sigma_k^-(s)\right]\sigma_j^z(s)-\nonumber\\
\!\!\left[J_{jk}^\perp+2\eta_k^-(s)\sigma_j^+(s)\right]\sigma_k^z(s)\Bigg\}ds+c_{jk}\Delta_{kj}(0,t)
\label{sig+sig-soln}
\end{align}
where $\Delta_{jk}'(s,t)=e^{2i\int_s^t\!\Delta\! B_{jk}^{\rm eff}(\tau)d\tau}$, and $c_{jk}=\sigma_j^+\sigma_k^-(0)$ is the initial condition.
Substituting Eq.~\eqref{sig+sig-soln} into Eq.~\eqref{sigz}, we find that the terms $2\eta_j^+(s)\sigma_k^-(s)$
and $2\eta_k^-(s)\sigma_j^+(s)$ within the square parentheses of Eq.~\eqref{sig+sig-soln} are summed over,
and hence can be ignored, due to our self-averaging approximation. We then proceed with the same mean-field approximation
as in Section~\ref{sec:model}, this time replacing $\Delta_{jk}'(s,t)\rightarrow\langle\Delta_{jk}'(s,t)\rangle$,
which is again assumed to be a Gaussian random variable, such that
\begin{equation}
 \langle\Delta_{kj}'(s,t)\rangle=e^{-2i\delta\!B_{kj}(t-s)}e^{-4\int_0^{|t-s|}K_{kj}'(\tau)\left(|t-s|-\tau\right)d\tau}
\end{equation}
where
\begin{align}
 K_{kj}'(t-t')&=\langle\Delta\!B_{kj}^{\rm fluct}(t)\Delta\!B_{kj}^{\rm fluct}(t')\rangle\nonumber\\
&=K_{kj}(t-t')+2\Lambda(t-t').
\end{align}
Proceeding in this way, Eq.~\eqref{sigz} for the time evolution of $\sigma_k^z$ becomes,
\begin{align}
 \partial_t\sigma_k^z(t)=&\sum_{j\neq k}\Gamma_{jk}'\left[\sigma_j^z(t)-\sigma_k^z(t)\right]+\xi_k(t)+\eta_k(t)-\nonumber\\
&8\int_0^t\Big\{\eta_k^-(t)e^{2i\int_s^tB_k^{\rm eff}(\tau)d\tau}\eta_k^+(s)+\nonumber\\
&\qquad\eta_k^+(t)e^{-2i\int_s^tB_k^{\rm eff}(\tau)d\tau}\eta_k^-(s)\Big\}\sigma_k^z(s)ds\label{sigzagain}
\end{align}
where
\begin{equation}
 \xi_k(t)=4i\sum_{j\neq k}J_{jk}^\perp\left[c_{jk}\Delta_{kj}'(0,t)-c_{kj}\Delta_{jk}'(0,t)\right]
\end{equation}
and
\begin{equation}
 \eta_k(t)=4i\left[\eta_k^+e^{-2i\int_0^tB_k^{\rm eff}(\tau)d\tau}c^--\eta_k^-e^{2i\int_0^tB_k^{\rm eff}(\tau)d\tau}c^+\right]
\end{equation}
are both noise terms, arising from the initial conditions of $\sigma_j^+\sigma_k^-$ and $\sigma_j^\pm$ respectively. The
new rate, $\Gamma'_{jk}$ is now given by
\begin{align}
 \Gamma_{jk}'=&4(J_{jk}^\perp)^2\int_{-\infty}^{\infty}\exp\!\!\Bigg[2i\delta\! B_{jk}s-\nonumber\\
&\qquad
\left.4\int_0^{|s|}K_{kj}'(\mu)\left(|s|-\mu\right)d\mu\right]ds,\label{Gammaprime}
\end{align}
where we have employed the Markov approximation, to remove $\left[\sigma_j^z(t)-\sigma_k^z(t)\right]$ from the
integral, and used the quickly decaying property of $\langle\Delta_{jk}'(s,t)\rangle$ to extend the
upper and lower limits of the integral to $\pm\infty$.

The integral term in Eq.~\eqref{sigzagain} can be greatly simplified by replacing the terms in the
curly parentheses by their average value. This approximation is consistent with an assumption of the
differing time scales between fluctuating local magnetic fields at site $k$, and the individual
dynamics of a single spin at site $k$. When a large number of individual spins contribute to the
local effective field $B_k^{\rm eff}$ at site $k$ (as is the case for systems with long range
interactions or high dimensionality) the fluctuations will appear Gaussian, and the term in
Eq.~\eqref{sigzagain} involving the integral, becomes
\begin{align}
 -8\int_0^t\Lambda(t-s)\cos&\left[2B_k(t-s)\right]\times\nonumber\\
&e^{-4\int_0^{|t-s|}G_k(\tau)\left[|t-s|-\tau\right]d\tau}\sigma_k^z(s)ds\label{integralterm}
\end{align}
where
\begin{align}
 G_k(\tau)=&\sum_{m\neq k}\sum_{n\neq k}J_{mk}^\parallel J_{nk}^\parallel\langle\sigma_m^z(0)\sigma_n^z(\tau)\rangle+
\Lambda(\tau).
\end{align}
The term preceeding $\sigma_k^z(s)$ in Eq.~\eqref{integralterm} decays much faster than the
evolution of $\sigma_k^z(s)$, so we can apply the Markov approximation $\sigma_k^z(s)\rightarrow\sigma_k^z(t)$,
and extending the upper and lower limits of integration to $\pm\infty$ we find
\begin{equation}\label{sigz4}
 \partial_t\langle\sigma_k^z\rangle=\sum_{j\neq k}\Gamma_{jk}'\langle\sigma_j^z-\sigma_k^z\rangle\!
-\! \Upsilon_k\langle\sigma_k^z\rangle
+\xi_k+\eta_k
\end{equation}
where
\begin{equation}
 \Upsilon_k=4\int_{-\infty}^\infty\Lambda(s)\cos\left(2B_ks\right)e^{-4\int_0^{|s|}G_k(\mu)\left[|s|-\mu\right]d\mu}ds
\end{equation}
gives a new rate at which the spin direction at site $k$ relaxes down into a completely random orientation of either $\pm1$.
{This} relaxation mechanism is entirely due to the fluctuating external magnetic field terms;
$\eta^\alpha_i(t)$, in {the Hamiltonian of Eq.~\eqref{hamnoise}.}

\subsection{Example: white noise}

If we {consider} the following simple example
\begin{equation}
 \Lambda(t)=\lambda\,\delta(t)
\end{equation}
then we find
 $\langle\eta_j(t)\eta_k(t')\rangle=8\delta_{jk}\lambda\,\delta(t-t')$
and
 $\Upsilon_k=4\lambda$.
We wish to examine two different limiting cases;
\begin{enumerate}
 \item $\sqrt{\lambda}\ll J_{ik}$ and $J_{ik}\ll B_i-B_k$
 \item $J_{ik}\ll \sqrt{\lambda}$ and $J_{ik}\ll B_i-B_k$
\end{enumerate}

\begin{figure}
\includegraphics[width=8cm]{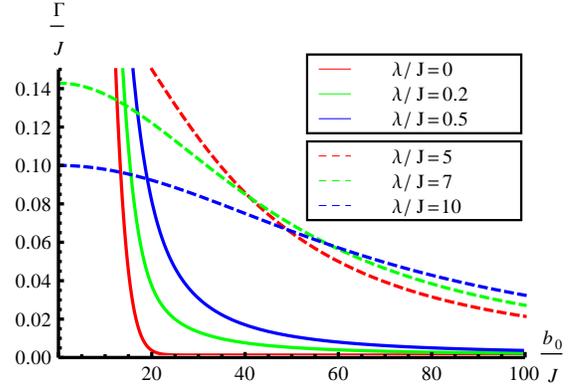}
\caption{Plotting the diffusion rate $\Gamma_{jk}'/J$ as a function of magnetic field gradient $(B_j-B_k)/J$ for a variety of
different values of $\lambda/J$. {The solid lines show the result in Eq.~\eqref{case1} for case 1. 
The dashed lines show the result of Eq.~\eqref{lorentzian} for case 2.}  The actual model is taken to be the same as the
Heisenberg model discussed in Section~\ref{heisenberg}. }\label{fig2}
\end{figure}

In {\bf case 1.} the external noise is sufficiently weak that the relaxation time-scale is essentially infinite,
in which case we can set $\langle\sigma_m^z(t)\sigma_n^z(t')\rangle\simeq\delta_{mn}$. In this way we find
$K'_{jk}(t)=\kappa_{jk}+2\lambda\delta(t)$, where
\begin{equation}
\kappa_{jk}=\sum_{m\neq\{j,k\}}\left(J_{mk}^\parallel-J_{mj}^\parallel\right)^2.
\end{equation}
{Continuing with the calculation}, we find the following expression for the rate;
\begin{equation}
 \Gamma_{jk}'=8(J_{jk}^\perp)^2\int_0^\infty ds\cos\left[2(B_k-B_j)s\right]e^{-2[\kappa_{jk}s^2+4\lambda s]}.
\end{equation}
This integral can be 
expanded to first order in the small parameter,
to give
\begin{align}\nonumber
 \Gamma_{jk}'&\simeq8(J_{jk}^\perp)^2\!\int_0^\infty \!\!ds\cos\left[2\delta\!B_{jk}s\right]e^{-2\kappa_{jk}s^2}\left(1-8\lambda s\right)
\\&=4(J_{jk}^\perp)^2\Biggl[\frac{\sqrt{\pi}\exp\left({-\frac{(\delta\! B_{jk})^2}{2\kappa_{jk}}}\right)}{\sqrt{2\kappa_{jk}}}-
\frac{4\lambda}{\kappa_{jk}}+\nonumber\\
&\qquad\qquad\qquad\qquad
\frac{4\sqrt{2}\delta\! B_{jk}\lambda F_{\rm D}\left(
\frac{\delta\! B_{jk}}{\sqrt{2\kappa_{jk}}}\right)}{\kappa_{jk}^{3/2}}\Biggr]\label{case1}
\end{align}
where $F_{\rm D}(x)=e^{-x^2}\int_0^x e^{y^2}dy$ is Dawsons integral~\cite{abramowitz}. 
{This result is shown in the solid lines of Fig.~\ref{fig2} for the case of the Heisenberg model 
on a cubic lattice (as discussed in Section~\ref{heisenberg}).} 
We can further approximate Dawsons integral, in the case of a large gradient $\delta\! B_{jk}\gg\sqrt{2\kappa_{jk}}$,
to give $F_{\rm D}(x)\simeq\frac{1}{2x}+\frac{1}{4x^3}+\O(x^{-5})$, for large $|x|$. From this we find the
asymptotic {behaviour} of the rate
\begin{equation}
 \Gamma_{jk}'\rightarrow\frac{16(J_{jk}^\perp)^2\lambda}{\delta\!B_{jk}^2},
\end{equation}
{valid when $\delta\!B_{jk}\gg J_{jk}^\perp$.}

In {\bf case 2.} the external noise is sufficiently strong, that it dominates over the interaction-induced
spin-diffusion process{. We can then} approximate Eq.~\eqref{sigz4} as
\begin{equation}
 \partial_t\langle\sigma_k^z\rangle\simeq- \Upsilon_k\langle\sigma_k^z\rangle
+\eta_k.
\end{equation}
In this case one would observe exponential decay in the autocorrelation function (due to
the noise term $\eta_k$) given by
\begin{equation}
 \langle\sigma_k^z(t)\sigma_j^z(t')\rangle=\delta_{jk}e^{-4\lambda |t-t'|}.
\end{equation}
This leads to $K_{jk}'(t)=2\lambda\delta(t)+\kappa_{jk}e^{-4\lambda|t|}\simeq2\lambda\delta(t)$, which
gives us the following expression for the rate;
\begin{equation}\label{lorentzian}
 \Gamma_{jk}'=16(J_{jk}^\perp)^2\frac{\lambda}{\delta\!B_{jk}^2+16\lambda^2}.
\end{equation}
{This result is shown in the dashed lines of Fig.~\ref{fig2} for the case of the Heisenberg model 
on a cubic lattice (as discussed in Section~\ref{heisenberg}).} 

Thus, in both cases 1. and 2. we find that the rate now decays as the inverse of the gradient squared;
$\sim\left(J_{jk}^\perp/\delta\!B_{jk}\right)^2$. This provides a huge contrast with the noiseless
situation of {Section}~\ref{sec:model}, where the rate decays as
$\sim\exp\left[-\left(J_{jk}^\perp/\delta\!B_{jk}\right)^2\right]$.
The presence of the noise provides a means for spin diffusion {to occur} over a much faster time-scale
{(in the presence of a strong external magnetic field gradient)}.

\section{Discussion and summary}\label{sec:discussion}

In Section~\ref{sec:model} of this article we have derived a dynamical mean-field theory for systems of spin-half particles
on a lattice{,} in the presence of
a nonuniform, external magnetic field. The theory is applicable in the case where the magnetic field gradient between two
lattice sites is large compared to the interactions. Additionally, the number of interacting pairs should be large (as is the
case for systems of high dimensionality or long range interactions). This condition is necessary to ensure the
fluctuations of the {\it effective} field at each lattice site are Gaussian (the central limit theorem).
One of the most notable approximations we made in deriving this theory of spin diffusion was the
exclusion of the summation in Eq.~\eqref{eq:sigmapm}. With this sum excluded, we were able to derive a
solution to Eq.~\eqref{eq:sigmapm}, shown in Eq.~\eqref{eq:sigmasol}. We can use this expression for
$\sigma_j^+\sigma_k^-(t)$, to estimate the size of the summation term in Eq.~\eqref{eq:sigmapm}, and thus
estimate the error in this approximation.

First, we note from Eqs.~\eqref{eq:sigsol2} and \eqref{Gamma}, the size of $\sigma_j^+\sigma_k^-(t)$
is roughly $\Gamma_{jk}/(2J_{jk}^\perp)$. Thus, if we substitute our expression for
$\sigma_i^+\sigma_k^-$ back into Eq.~\eqref{eq:sigmapm}, we see that the size of the summation term is
approximately $\max\left\{\Gamma_{ni},\Gamma_{nk}\right\}$ where $n$ runs over lattice sites which
are {\it mutual} neighbors of sites $i$ and $k$. Assuming a certain level of isotropy exists within the system,
we conclude that, provided $J_{ik}^\perp\gg\Gamma_{ik}$, for all interacting pairs $i$ and $k$, the
exclusion of the summation in Eq.~\eqref{eq:sigmapm} is justified.
With all conditions satisfied, the equation of motion
for the $z$-component of the individual spins is a Langevin equation with additive noise, see Eqs.~\eqref{sigz3}.

If the condition $J_{ik}^\perp\gg\Gamma_{ik}$ were not satisfied, and the summation in Eq.~\eqref{eq:sigmapm} could not
be justifiably ignored, we would expect a similar analysis to be possible. The summation term would manifest as
multiplicative noise in the coefficients $\Gamma_{ik}$ of the Langevin equation~\eqref{sigz3}, as well as the
additive noise which we have derived. Further work on this issue however, is still in progress, and the details deferred to a future
publication.

The model can be described in terms of simple physical principles, as illustrated in Fig.~\ref{fig3}. Interactions between sites $i$ and $j$ can cause spin
\emph{flip-flopping}, i.e. $|i_{\uparrow}j_{\downarrow}\rangle\rightleftarrows |i_{\downarrow}j_{\uparrow}\rangle$.
This process occurs when sites $i$ and $j$ have opposite spin, and does not conserve energy when the external
field gradient is nonzero (due to the different Zeeman energies). These sites $i$ and $j$ however, also interact with all
other neighboring lattice sites (the number of which is assumed to be large).
A crucial approximation in our model is to treat all remaining sites as composing
an effective bath, or rapidly fluctuating environment in which sites $i$ and $j$ inhabit [see Fig.~\ref{fig3} (b)].
In this way, one can derive the rate at which the spin \emph{flip-flopping} occurs (we have labelled this quantity
$\Gamma_{ij}$), and naturally it will
depend on the bath parameters. To be more specific, it depends on the correlation functions between neighboring sites within
the bath.
The final step then is to determine the rate $\Gamma_{ij}$ that is self-consistent with the bath, i.e. the value of
$\Gamma_{ij}$ which yields the {\it same correlation function} between neighboring sites, as that from which it was derived.


We find the rate $\Gamma_{ij}$ decays very quickly with increasing field gradient. Equation~\eqref{Gamma1} predicts the rate
decays in the same way as a Gaussian distribution. From a numerical study of the cubic Heisenberg lattice (presented in
Section~\ref{heisenberg}), we expect this prediction to be
accurate for $B_i-B_j\gtrsim10J_{ij}$ (see Fig.~\ref{fig1}).  This result implies that the observation of spin diffusion in
systems with a very strong magnetic field gradient is likely to be difficult as the diffusion time-scales would be
very large.

However, in {Section}~\ref{relaxation} we studied the influence {of} external noise on this rate{. The 
presence of the external noise} 
turns out to be favourable for increasing the diffusion rates. We made use of the same set of assumptions
in deriving a second Langevin equation [see Eq.~\eqref{sigz4}]. In contrast to {Section}~\ref{sec:model}, the Langevin equation
now includes a decay-constant, denoted $\Upsilon_k$, which relaxes the system down into {a} state where the orientation of
the magnetic moment is completely random, i.e. $\langle\sigma_k^z\rangle=0$.
Spin flip-flops still occur in the system, and the rate at which they occur; $\Gamma_{ij}'$, is affected by the noise.
As a general rule, the rate $\Gamma_{ij}'$ increases with increasing noise, as is illustrated in Fig.~\ref{fig2}.
In the limiting case where the external noise is far greater than both the interaction coupling and the
external field gradient, we find the rate $\Gamma_{ij}$ decays in the same way as a Cauchy-Lorentz distribution,
see Eq.~\eqref{lorentzian}.
This predicted increase in the rate may help to explain experiments where diffusion has purportedly been observed
in systems with very large magnetic field gradients.

\section{Acknowledgements}
We thank Olexander Chumak, Chris Hammel and Semion Saykin for valuable discussions. 
The work is supported by the US DOE, and, in part, by ONR and NAS/LPS. Andrew Sykes gratefully 
acknowledges the support of the U.S. Department of Energy through the LANL/LDRD Program for this 
work.

\begin{appendix}

\section{Perturbation theory for the two-body problem in a fluctuating external field}\label{appA}
Consider the following time dependent Hamiltonian describing two spin-half particles located at sites 1 and 2,
interacting via an exchange interaction,
\begin{equation}
 \hat{H}=B_1(t)\sigma_1^z+B_2(t)\sigma_2^z+J\left(\sigma_1^x\sigma_2^x+\sigma_1^y\sigma_2^y+\sigma_1^z\sigma_2^z\right)
\end{equation}
where the external magnetic field $B_i(t)=B_i+b_i(t)$ consists of
a constant part and a fluctuating part.
We wish to calculate the probability of the spins {\it flip-flopping} in time $t$, that is
\begin{equation}\label{p}
 p(t)=\Big|\langle 1_{\downarrow}2_{\uparrow}|\hat{U}(t,0)|1_{\uparrow}2_{\downarrow}\rangle\Big|^2
\end{equation}
where $\hat{U}(t,0)$ is the time evolution operator for $\hat{H}$. Splitting the full Hamiltonian up into
a noninteracting and an interacting part,
\begin{align}
 \hat{H}_0&=B_1(t)\sigma_1^z+B_2(t)\sigma_2^z\\
 \hat{V}&=J\left(\sigma_1^x\sigma_2^x+\sigma_1^y\sigma_2^y+\sigma_1^z\sigma_2^z\right)
\end{align}
and moving to the interaction picture; $|\Psi_{\rm S}(t)\rangle=\hat{U}_0(t,0)|\Psi_{\rm I}(t)\rangle$
where $\hat{U}_0(t,0)=\exp\left[-i\int_{0}^t\hat{H}_0(\tau)d\tau\right]$ is the time evolution
operator of the noninteracting Hamiltonian. Defining $\hat{V}_I(t)=\hat{U}_0(t,t_0)\hat{V}\hat{U}_0(t_0,t)$
and $\hat{U}_I(t,0)\exp_+\left[-i\int_{0}^t\hat{V}_I(\tau)d\tau\right]$, where $\exp_+$ denotes the usual
time-ordered Dyson series~\cite{sakurai} (appropriate for noncommuting $[\hat{V}_I(t_1),\hat{V}_I(t_2)]\neq0$ when $t_1\neq t_2$).

Using this standard formalism, we approximate the full time evolution operator as,
\begin{align}
 \hat{U}(t,0)&=\hat{U}_0(t,0)\hat{U}_I(t,0)\nonumber\\
&\simeq\hat{U}_0(t,0)\left[1-i\int_{0}^t\hat{V}_I(\tau)d\tau\right]
\end{align}
by truncating the Dyson series for $\hat{U}_I$. Substituting this approximation into
Eq.~\eqref{p} and working through the calculation in a straight-forward manner we
arrive at
\begin{align}\nonumber
 p(t)=4 J^2\int_0^t d\tau_1\int_0^t d\tau_2\,e^{-2i(B_1-B_2)(\tau_1-\tau_2)+2i\int_{\tau_1}^{\tau_2}\delta b(s)ds}
\end{align}
where $\delta b(s)=b_1(s)-b_2(s)$.
At this point it is convenient to take an average over the fluctuating component of the external
field, $e^{2i\int_{\tau_1}^{\tau_2}\delta b(s)ds}\rightarrow\langle e^{2i\int_{\tau_1}^{\tau_2}\delta b(s)ds}\rangle$.
Assuming these fluctuations are Gaussian, and time-translationally invariant, we find
\begin{align}\nonumber
 p(t)=\,&4 J^2\int_0^t d\tau_1\int_0^t d\tau_2\,e^{-2i(B_1-B_2)(\tau_1-\tau_2)}\times\\
&\quad e^{-4\int_{0}^{|\tau_1-\tau_2|}ds\langle\delta b(0)\delta b(s)\rangle\left(|\tau_1-\tau_2|-s\right)}\nonumber\\
=\,& 4J^2\int_{-t}^t d\tau\,\left(t-|\tau|\right) e^{-2i(B_1-B_2)\tau}\times\nonumber\\
&\quad e^{-4\int_{0}^{|\tau|}ds\langle\delta b(0)\delta b(s)\rangle\left(|\tau|-s\right)}.
\end{align}
In the limit then, where the time $t$ is much larger than the time scale over which the final term
$e^{-4\int_{0}^{|\tau|}ds\langle\delta b(0)\delta b(s)\rangle\left(|\tau|-s\right)}$ decays,
the probability becomes
\begin{align}
 p(t)=\,&4tJ^2\int_{-\infty}^\infty \exp\!\!\Bigg[2i(B_1-B_2)s-\nonumber\\
&\qquad
\left.4\int_0^{|s|}\langle\delta b(0)\delta b(s)\rangle \left(|s|-\mu\right)d\mu\right]ds\label{probability}
\end{align}
This probability in Eq.~\eqref{probability} should be compared to the rate at which spin flips are predicted to
occur from Eq.~\eqref{Gamma} in Section~\ref{sec:model}.
In making this comparison, we see that the approximations we have applied in deriving the equation of
motion~\eqref{sigz3} for $\sigma_k^z$ amount to treating all sites other $j$ and $k$ as composing
an effective bath (equivalent to a fluctuating external field).

\section{Numerical algorithm for solving the integral equation}\label{appB}

For our particularly simple choice of $B({\bf r})=\frac{b_0}{\sqrt{3}}(r_x+r_y+r_z)$, the integral equation
we must solve is simply Eq.~\eqref{Gamma2} with $K(\mu)$ given by Eq.~\eqref{f}. From this equation, we wish
to determine $\Gamma$ as a function of $b_0$, the dependence on $J$ can be removed, by switching to variables
$\tilde{\Gamma}=\Gamma/J$ and $\tilde{b}_0=b_0/J$, such that we have
\begin{equation}\nonumber
\tilde{\Gamma} = 4\int_{-\infty}^\infty ds \,e^{\frac{2i\tilde{b}_0 s}{\sqrt{3}}}
e^{-8\int_0^{|s|}d\mu\, e^{-6\tilde{\Gamma}\mu}f(2\tilde{\Gamma}\mu)\left(|s|-\mu\right)}.
\end{equation}
We then search for a root of this equation, by iterating
\begin{equation}\label{GammaNumerics1}
\tilde{\Gamma}^{(n+1)} = 4\int_{-\infty}^\infty \!\!\! ds \,e^{\frac{2i\tilde{b}_0 s}{\sqrt{3}}}
e^{-8\int_0^{|s|}d\mu\, e^{-6\tilde{\Gamma}^{(n)}\mu}f(2\tilde{\Gamma}^{(n)}\mu)\left(|s|-\mu\right)}.
\end{equation}
for $n=0,1,2,\ldots$ up to convergence, which in our case was chosen to be
$|\tilde{\Gamma}^{(n+1)}-\tilde{\Gamma}^{(n)}|<10^{-4}$. In order to choose a reasonable
initial prediction for $\tilde{\Gamma}^{(0)}$ we begin the algorithm at $\tilde{b}_0=20$, and
define
\begin{equation}\label{Gamma0}
\tilde{\Gamma}^{(0)} \simeq 4\sqrt{\frac{\pi}{20}}e^{-\tilde{b}_0^2/(60)}.
\end{equation}
Once the algorithm has converged, we decrease $b_0$ by a small amount and use our previous
prediction for $\tilde{\Gamma}$ as our new $\tilde{\Gamma}^{(0)}$.

\section{The issue regarding convergence/divergence of $\Gamma$ as $b_0\rightarrow0$}\label{appc}

As the field gradient decreases in a particular direction, the rate at which spin {\it flip-flops} occur in that particular
direction increases, see Figure~\ref{fig1}. It is not clear, a priori, that the rate will remain finite in the limit of
vanishing gradient. Consider, for example, the RHS of Eq.~\eqref{Gamma2} (and set $J=1$). We can rewrite this in terms of the
Fourier transform of $K(\mu)=\frac{1}{2\pi}\int e^{-i\omega\mu}\tilde{K}(\omega)d\omega$, and we are only interested in the
case where $b_0=0$, so we find,
\begin{equation}
 {\rm RHS}(\Gamma)=4\int_{\mathbb{R}} \exp\left[-\frac{2}{\pi}\int_{\mathbb{R}}
\tilde{K}(\omega)\frac{1-\cos(\omega|s|)}{\omega^2}d\omega\right]ds.
\end{equation}
Next we define $\gamma=\mu\Gamma$, in which case
\begin{align}
\tilde{K}(\omega)&=2\int_{\mathbb{R}}\frac{d\gamma}{\Gamma} e^{i\frac{\omega\gamma}{\Gamma}}e^{-6\gamma}f(2\gamma)\nonumber\\
&=\frac{2}{\Gamma}H\left(\frac{\omega}{\Gamma}\right)
\end{align}
where $H(x)=\int_{\mathbb{R}}d\gamma\,e^{ix\gamma}e^{-6\gamma}f(2\gamma)$. The RHS therefore becomes,
\begin{equation}
 {\rm RHS}(\Gamma)=4\int_{\mathbb{R}} \exp\left[-\frac{4}{\pi}\int_{\mathbb{R}}d\eta\,
H(\eta)\frac{1-\cos(\Gamma\eta|s|)}{\Gamma^2\eta^2}\right]ds.
\end{equation}
where we defined $\eta=\omega/\Gamma$. Now we make the assumption that $\Gamma$ does become very large, in this limit we find
\begin{equation}
 \frac{1-\cos(\Gamma\eta|s|)}{\Gamma^2\eta^2}\rightarrow\frac{\pi|s|}{\Gamma}\delta(\eta)
\end{equation}
and therefore
\begin{equation}
 {\rm RHS}(\Gamma)\rightarrow \frac{2\Gamma}{H(0)}
\end{equation}
for large $\Gamma$. The quantity $H(0)$ can be calculated numerically to be $H(0)\eqsim2.33$, thereby indicating that the slope
of the RHS is $<1$ for large $\Gamma$. From this we conclude that a finite value of $\Gamma$ will exist at $b_0=0$ which
satisfies Eq.~\eqref{Gamma2}.

\end{appendix}


\end{document}